\documentclass[sigconf,natbib=false,pbalance=true]{acmart}
\usepackage{subcaption}
\usepackage{graphicx}
\usepackage{wrapfig}
\usepackage{subcaption}
\usepackage{enumitem}
\usepackage{booktabs}
\usepackage{placeins}
\usepackage{tabularx}
\usepackage{array}
\usepackage{amsmath}
\usepackage{algorithmic}
\usepackage{graphicx}
\AtBeginDocument{%
  \providecommand\BibTeX{{%
    \normalfont B\kern-0.5em{\scshape i\kern-0.25em b}\kern-0.8em\TeX}}}

\usepackage{textcomp}
\usepackage{xcolor}
\usepackage{placeins}
\usepackage{multirow}
\usepackage[ruled,vlined]{algorithm2e}
\usepackage[symbol,flushmargin]{footmisc}
\usepackage{soul}
\usepackage{subcaption}
\usepackage{booktabs}
\usepackage{graphicx} % for \resizebox
\usepackage{siunitx}  % for S columns
\usepackage{makecell}
\usepackage{multirow}
\newcolumntype{Y}{>{\centering\arraybackslash}X}
\usepackage{soul}
\usepackage{xcolor}
\sethlcolor{red!20}

\setcopyright{none}
\renewcommand\footnotetextcopyrightpermission[1]{}

\begin{document}

%%
%% The "title" command has an optional parameter,
%% allowing the author to define a "short title" to be used in page headers.
\title{Application Agnostic EM Side-Channel Emanations of the FPGA Clock Distribution Network}

%%
%% The "author" command and its associated commands are used to define
%% the authors and their affiliations.
%% Of note is the shared affiliation of the first two authors, and the
%% "authornote" and "authornotemark" commands
%% used to denote shared contribution to the research.
\author{Ashish Sharma}
\email{as5463@drexel.edu}
\affiliation{%
  \institution{Drexel University}
  \city{Philadelphia}
  \state{Pennsylvania}
  \country{USA}
}

\author{Alia Long}
\email{alia.long@inl.gov}
\affiliation{%
  \institution{Idaho National Laboratory}
  \city{Idaho Falls}
  \state{Idaho}
  \country{USA}
}

\author{Keith D. Mecham}
\email{keith.mecham@inl.gov}
\affiliation{%
  \institution{Idaho National Laboratory}
  \city{Idaho Falls}
  \state{Idaho}
  \country{USA}
}

\author{Ioannis Savidis}
\email{is338@drexel.edu}
\affiliation{%
  \institution{Drexel University}
  \city{Philadelphia}
  \state{Pennsylvania}
  \country{USA}
}

%%
%% By default, the full list of authors will be used in the page
%% headers. Often, this list is too long, and will overlap
%% other information printed in the page headers. This command allows
%% the author to define a more concise list
%% of authors' names for this purpose.
%\renewcommand{\shortauthors}{Trovato et al.}

%%
%% The abstract is a short summary of the work to be presented in the
%% article.
\begin{abstract}
Electromagnetic side channels have been extensively utilized in non-invasive attacks or analyses to extract critical information from deployed systems. Most of the attacks or analyses are typically executed on cryptographic algorithms under the assumption that the underlying implementation remains fixed. The assumption is valid in the context of fixed hardware, where modifications to the circuit are not made after manufacturing. However, in the case of adaptive hardware, including field-programmable gate arrays (FPGAs), the assumption no longer holds as modifications to the configuration of the array are permitted post-manufacturing. Consequently, this study introduces a methodology for the analysis of electromagnetic side channels of an FPGA through the study of deterministic finite state machines (DFSMs), where the effects of placement, routing, and operating frequency on the EM emanations of the FPGA due to the utilized clocking resources are characterized. To analyze the impact of frequency, device placement, and resource allocation on the EM emanations from the FPGA, statistical analysis including mean difference and variance difference calculations of measured EM fields were utilized to identify points of high EM activity on a set grid of sampled points. A 1.32x increase in flagged grid points was observed in the analysis of the mean difference when the percentage of utilized clock resources was increased from 25\% to 90\%. Similarly, a 1.39x increase in flagged grid points was observed when the variance difference was calculated for the same increase in clock resources. 
%%%%%%%, such as differences in mean and variance, was% 
%% Add 2-3 lines to the abstract for summarizing the primary results

\end{abstract}

%%
%% The code below is generated by the tool at http://dl.acm.org/ccs.cfm.
%% Please copy and paste the code instead of the example below.
%%

%%
%% Keywords. The author(s) should pick words that accurately describe
%% the work being presented. Separate the keywords with commas.
\keywords{FPGA, EM, Clock Distribution Network, Side-Channels}

%%
%% This command processes the author and affiliation and title
%% information and builds the first part of the formatted document.
\maketitle

\thispagestyle{plain}
\pagestyle{plain}
\section{Introduction}
Over the past four decades, field-programmable gate arrays (FPGAs) have evolved from specialized device to prototype circuits into widely used computing platforms. FPGAs offer the ability to optimize and reconfigure implemented circuits numerous times after silicon fabrication, while 
%%%Need Reference for examples P_fpga>P_ASICs
reducing time-to-market and lowering operational costs. The unique advantages provided by FPGAs, relative to application-specific integrated circuits (ASICs), has led to increased adoption across various sectors, including cloud computing, implementation of algorithms for artificial intelligence, and circuits programmed for aerospace and defense systems. Such cross-sector utilization underscores the importance of safeguarding FPGA devices, particularly from side-channel attacks, which pose considerable threats to the security of the hardware platform. Diverse categories of side channels, which include power \cite{citezhao, Jaya_instruct}, electromagnetic (EM) \cite{citewerner1, Wang_SoC, Wang_AM}, temperature \cite{citehutter, zhang_temp}, and backscattered \cite{citenguyen} signals, enable the non-invasive extraction of critical information that conventional software-based methods are not able to achieve. 
\begin{figure}[h!]
  \centering
  \begin{subfigure}[t]{0.35\columnwidth}
    \centering
    \includegraphics[width=\linewidth]{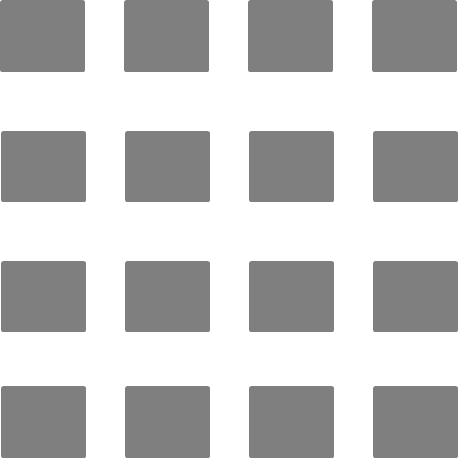}
    %\caption{Unconfigured}
    \caption{}
    \label{fig:a}
  \end{subfigure}\hfill
  \begin{subfigure}[t]{0.6\columnwidth}
    \centering
    \includegraphics[width=\linewidth]{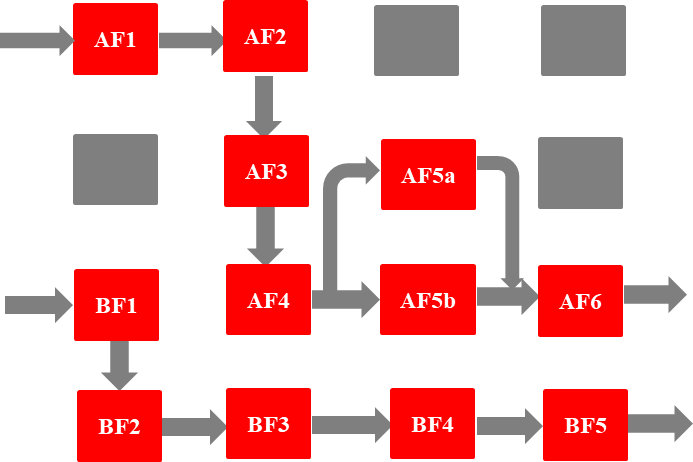}
    %\caption{Configured}
    \caption{}
    \label{fig:b}
  \end{subfigure}
  \caption{Representation of (a) unconfigured and (b) configured adaptive hardware platform \cite{citeadapt}.}
  \vspace{-8pt}
  \label{fig:adept}
\end{figure}

Electromagnetic (EM) side-channel characterization has been extensively utilized to conduct non-malicious analysis for the detection of counterfeit components or hardware Trojans ~\cite{trojan}. Understanding the sources of electromagnetic emissions within integrated circuits (ICs) is, therefore, crucial, as such characterization enables the identification of system vulnerabilities and prevents adversaries from executing non-invasive attacks to extract sensitive information.

For FPGAs, EM signals are spread across two distinct spectral regions: the low-frequency range ($\approx$ 10 kHz to 1 MHz) and the high-frequency range ($ > $ tens of MHz). The primary emitters of low-frequency EM signals are large supply-current loops passing through power pins, PCB planes, and decoupling capacitors \cite{citewerner1}. In contrast, the primary emitters of high-frequency EM emanations are periodic signals including the primary on-chip clock, the SERDES reference clocks, and switching regulators. While low-frequency EM signals carry valuable exploitable information, high-frequency EM emanations modulated by the switching activity of logical devices generate data-dependent sidebands that extend through multiple harmonics, which offers a wider bandwidth of information and radiates more effectively. Such EM emanations, therefore, enable the non-invasive remote detection of confidential on-chip information.

\begin{figure*}[h!]
  \centering
  \begin{subfigure}[t]{0.24\textwidth}
    \centering    \includegraphics[width=\linewidth,height=15cm,keepaspectratio]{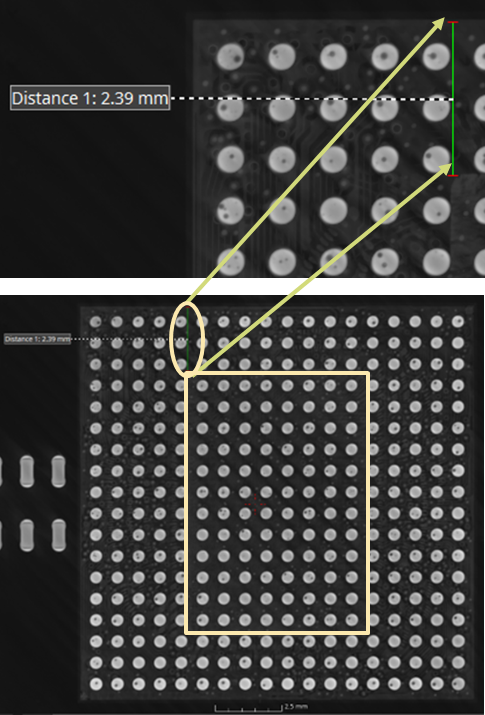}
    \caption{}
    \label{fig:}
  \end{subfigure}\hfill
  \begin{subfigure}[t]{0.24\textwidth}
    \centering    \includegraphics[width=\linewidth,height=6.65cm,keepaspectratio]{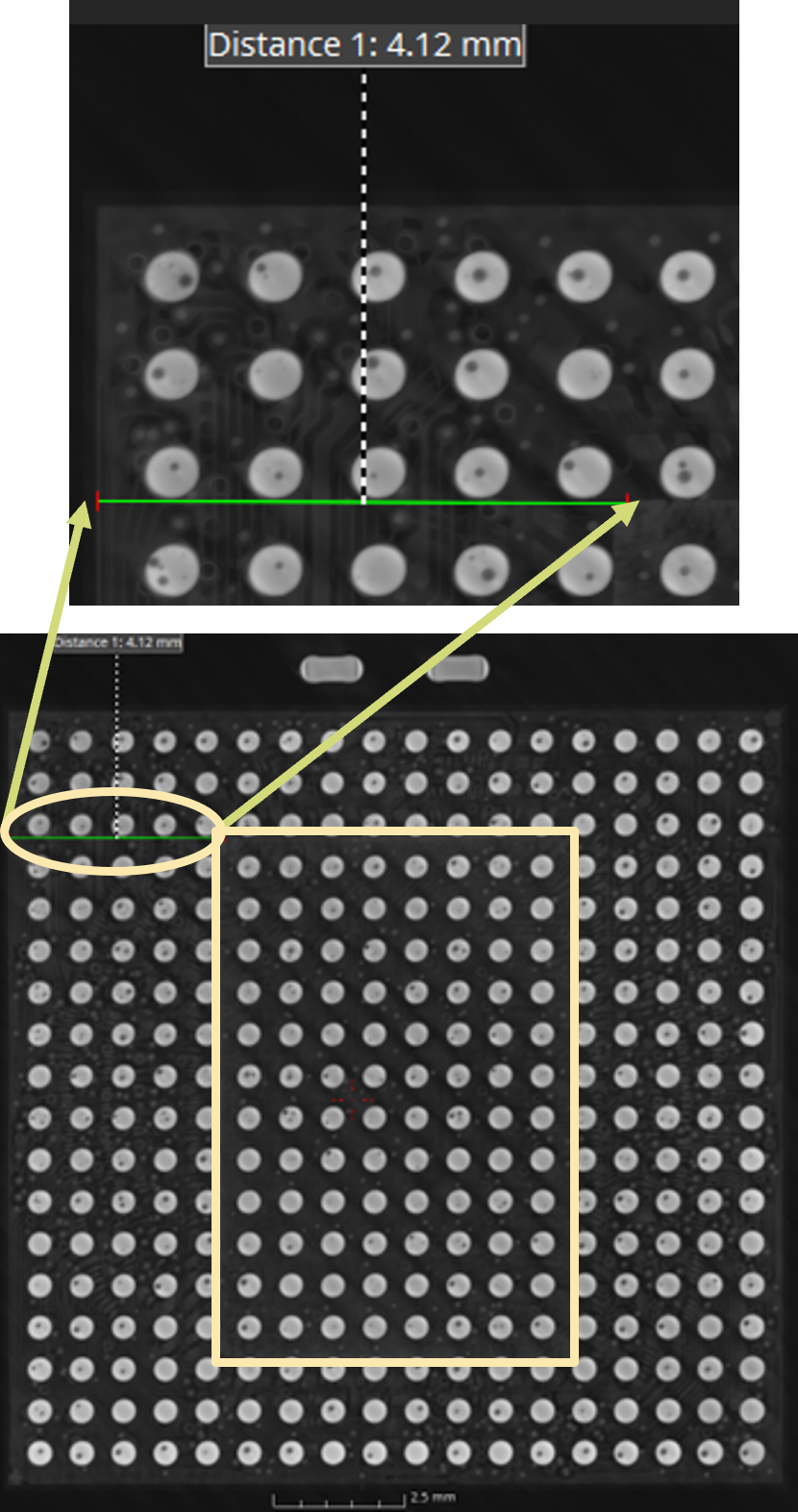}
    \caption{}
    \label{fig:row4_b}
  \end{subfigure}\hfill
  \begin{subfigure}[t]{0.24\textwidth}
    \centering    \includegraphics[width=\linewidth,height=6.65cm,keepaspectratio]{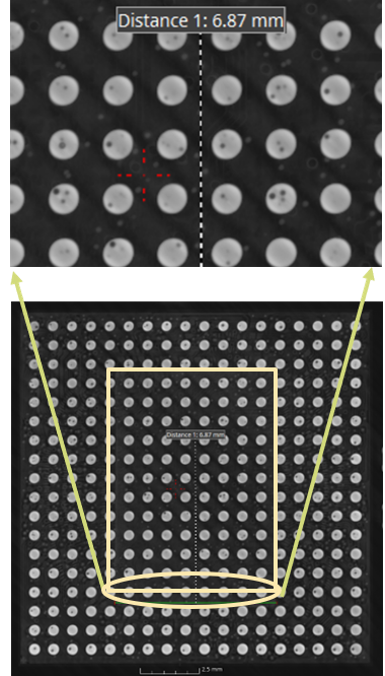}
    \caption{}
    \label{fig:row4_c}
  \end{subfigure}\hfill
  \begin{subfigure}[t]{0.24\textwidth}
    \centering    \includegraphics[width=\linewidth,height=6.65cm,keepaspectratio]{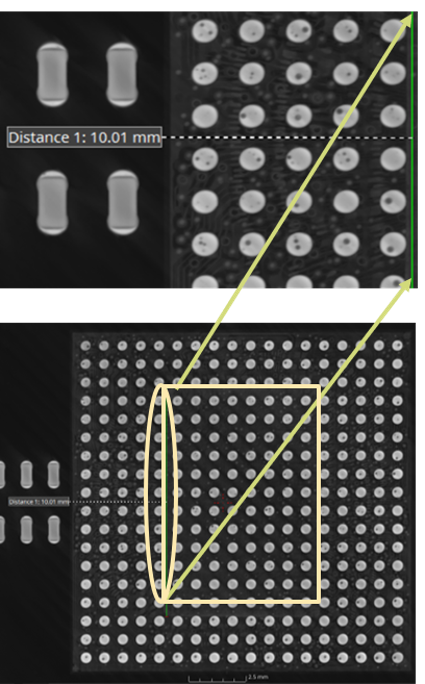}
    \caption{}
    \label{}
  \end{subfigure}

  \caption{Microphotographs of CT scanned Artix-7 depicting the (a) vertical distance between the top edge of the package and the top edge of the active die area, (b) horizontal distance between the left edge of the package and the left edge of the active die area, (c) width of the active die area, and (d) height of the active die area}
  \label{fig: CT_scan}
\end{figure*}

The method described in \cite{citewerner1} explores localization of the EM side-channel for board-level analysis, covering frequencies from several kilohertz to a gigahertz. For the analysis, a shielded probe that forms a 20 mm outer diameter loop \cite{wernerprobe} is positioned 3.5 cm above the device-under-test (DUT) to measure the H-field emanating from the PCB surface. A second probe is fixed in the same position on the DUT during each scan to provide a common reference needed to compare all measurement locations. The experimental setup described in \cite{citewerner1} is structured such that the electromagnetic emissions produced by the various functions implemented on the PCB are captured, instead of analyzing EM emanations from the active silicon area of the die.

Adaptive hardware, including field-programmable gate arrays (FPGAs), offer the capability of configuring standard circuit components within configurable logic blocks. In addition, configurable connectivity is provided through programmable interconnects optimized to meet the target performance requirements, an example of which is shown in Fig.~\ref{fig:adept}. Compared to field-programmable gate arrays (FPGAs), any code or cryptographic algorithm implemented on a central processing unit (CPU), graphics processing unit (GPU), or microcontroller is predominantly executed as a fixed set of instructions on a fixed hardware~\cite{Jaya_instruct}. In such configurations, the datapath and control logic are implemented as a predetermined microarchitecture. The algorithms are, therefore, compiled into instructions or kernels that are scheduled onto fixed execution units, which include the arithmetic logic unit (ALU), the single instruction multiple data (SIMD) unit, and caches. Consequently, optimization primarily targets the compiler, runtime scheduler, or memory management unit. 

Considering the fundamental distinction between fixed and configurable hardware, and the security risks posed by the leakage of high-frequency EM side-channel signals from an FPGA, this study presents a novel methodology to analyze the profile of the EM emanations originating from an FPGA with sub-millimeter (sub-mm) sampling. The developed approach is applied to analyze a deterministic finite state machine (DFSM), with particular emphasis on variations in input states, clock frequency, device placement, and utilized resources of the reconfigurable clock distribution network.

% Considering the fundamental distinctions between fixed and configurable hardware and the risks that high-frequency EM side-channel signals pose to FPGA security, this study presents a novel approach to analyze range through a comprehensive understanding of a deterministic finite state machine (DFSM), with particular emphasis on state, clock, and reset behavior within the reconfigurable clock distribution network. The approach is grounded in a detailed understanding of a deterministic finite state machine (DFSM). It focuses specifically on state, clock, and reset behavior within the reconfigurable clock distribution network.

The remainder of the paper is organized as follows: Background and a summary of prior research efforts are described in Section 2. The properties of an AMD Artix-7 FPGA and the area of the packaged FPGA that is scanned and characterized are discussed in Section 3. Details of the experimental setup utilized to measure electromagnetic (EM) emissions across various operating conditions are provided in Section 4. Analysis of the experimental results across the different operating points of the FPGA is presented in Section 5. Finally, some concluding remarks are provided in Section 6.  

\section{Background And Related Work}

%%%%%%%%%%%%%%% Required Edits %%%%%%%%%%%%%%%%%%%%%%% 
%% Add 2-3 sentences introducing the section and sub-sections

Prior methodologies developed to conduct electromagnetic (EM) side-channel analysis across fixed and adaptive hardware platforms are described. Non-invasive EM side-channel attacks based on probing and scanning are also analyzed, along with any limitations such techniques exhibit.

\begin{itemize}[leftmargin=0pt, label={}]
  \item \textbf{EM Side-Channel Analysis:} Electromagnetic side-channel analysis has been explored as a viable attack across computing platforms including fixed hardware devices such as the Arduino UNO \cite{citebaidual}, the 32-bit ARM microcontroller \cite{citescnif}, and the A13 OLinuXino \cite{citewerner1}, as well as for adaptive hardware such as the ALTERA Cyclone FPGA \cite{citecarlier}. The attack often targets the instruction sets of computing units or the implemented cryptographic algorithms including the advanced encryption standard (AES). In \cite{citecarlier}, a 128-bit AES algorithm was implemented on an ALTERA Cyclone FPGA operating at 50 MHz. The implemented circuit includes an explicit representation of the 128-bit internal state of the algorithm, which secures the entire plaintext during each executed round. During each round, the AES algorithm utilizes functions that include SubBytes, ShiftRows, MixColumns, and AddRoundKey. The resulting temporal features are analyzed in relation to the execution of the AES algorithm and a simple difference in the measured amplitude of the electromagnetic (EM) field intensity between different keys, which demonstrates consistency with Hamming-weight-related leakage analysis. Similarly, in \cite{citescnif}, a 128-bit AES algorithm was implemented on an ARM microcontroller, where a correlational electromagnetic analysis attack was executed by evaluating test vector leakage and the signal-to-noise ratio.   
     \begin{figure*}[h!]
  \centering
  \begin{subfigure}[t]{0.35\textwidth}
    \centering    \includegraphics[width=\textwidth,height=0.65\textheight,keepaspectratio]{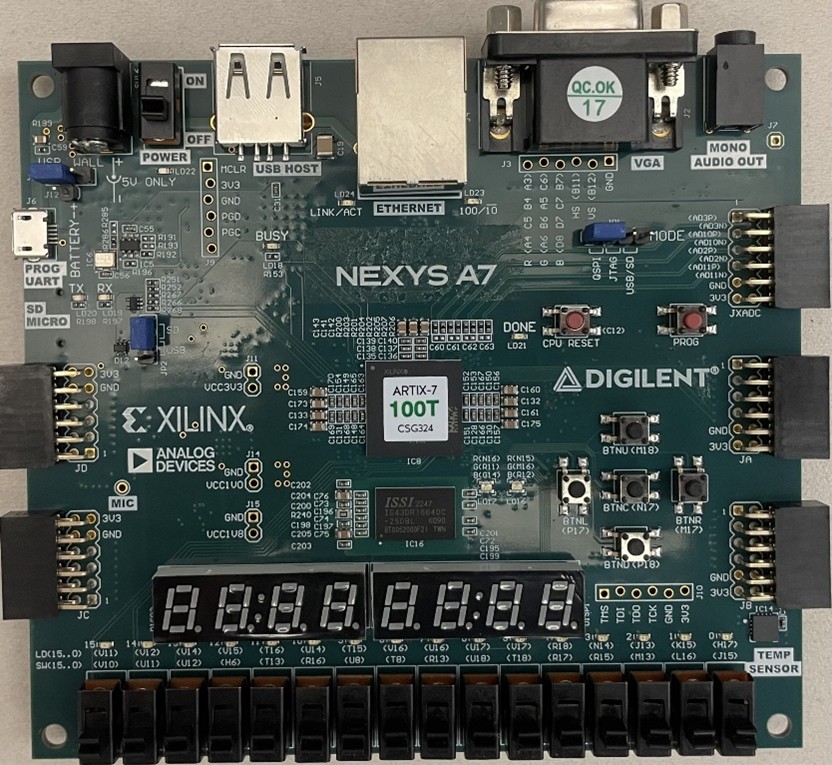}
    %\caption{Nexys A7.}
    \caption{}
    \label{fig:Nexys_A7}
  \end{subfigure}\hfill
  \begin{subfigure}[t]{0.60\textwidth}
    \centering    \includegraphics[width=\textwidth,height=0.65\textheight,keepaspectratio]{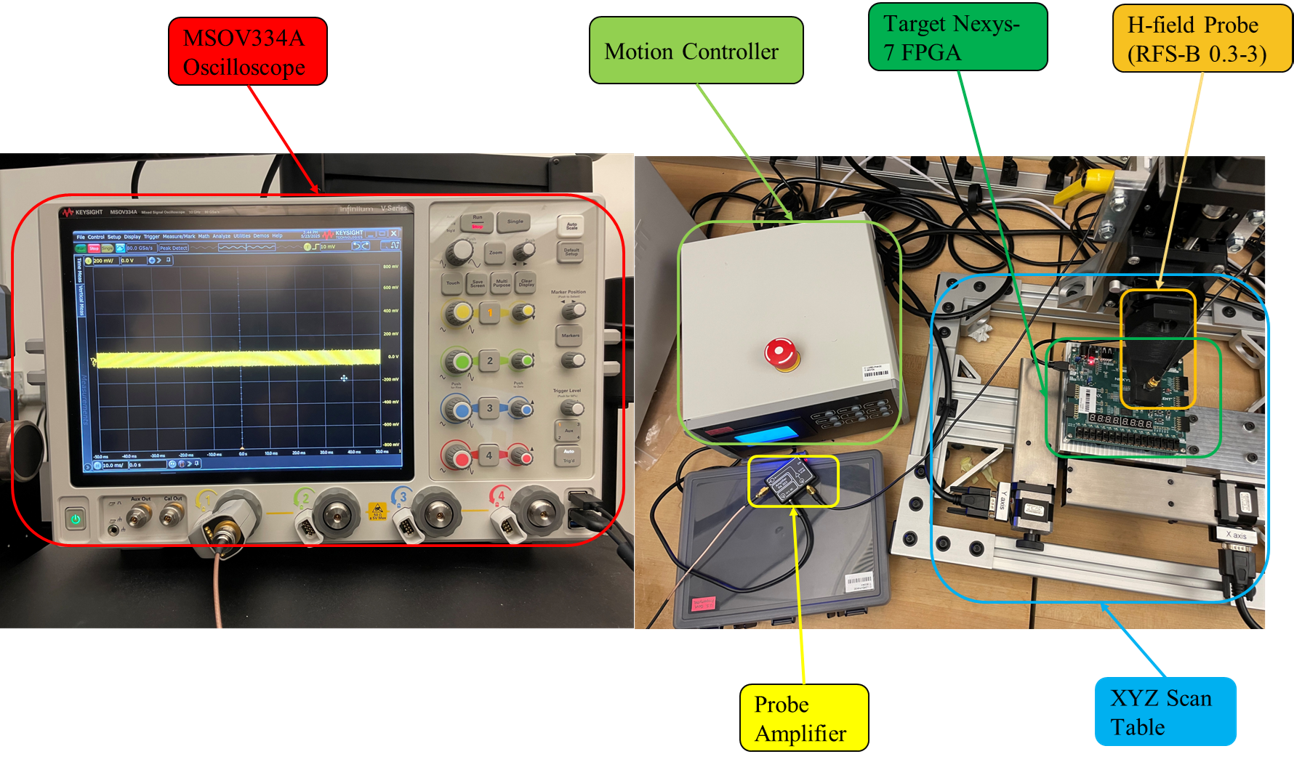}
    \caption{}
    %\caption{Components of EM scan station.}
    \label{fig:EM_SC_setup}
  \end{subfigure}
  \caption{Images of the (a) target Nexys A7 FPGA development board, and (b) the experimental configuration used for EM scanning and trace collection of the EM emanations of an Artix-7 FPGA. }
  \label{fig:Complete_EM_Setup}
\end{figure*}

  \item \textbf{Probe and Scan Methodology:} To perform a non-invasive attack or analysis, various probing and scanning methodologies have been developed. Methods include designing custom probes to evaluate leakage across the PCB \cite{citewerner1} or using commercially available probes \cite{Wang_SoC, citescnif} to develop a scanning procedure that identifies leakage points specifically for an implemented 128-bit AES algorithm.

  In \cite{citescnif}, a grid of points was defined across the entire accessible top surface of the integrated circuit for scanned measurement, which also included the physical package. A cost-effective scanner is used to collect traces at designated target points identified through assessment of the leaked EM field of an executing 128-bit AES core.  
  \item \textbf{Observation:} Most analyses of the electromagnetic (EM) emanations of a circuit are conducted as an attack, where a fixed implementation of a cryptographic algorithm, such as the 128-bit AES algorithm~\cite{Wang_SoC, Wang_AM}, is considered on a specific target device. The methods are limited by the difficulty of executing the attack, the type of analysis required, and the efficacy of the leakage measurements, as the implementation of the algorithm is associated with a particular micro-architecture and physical configuration. However, algorithms are implemented in various ways. For example, implementations of the 128-bit AES algorithm include iterative, fully unrolled, or pipelined configurations, composite S-box or table/ROM S-box configurations, different clock and control structures, and varying levels of parallelism. Specifically, in the case of adaptive hardware such as FPGAs, each configuration of an implemented circuit is affected by the algorithms executed to perform technology mapping, placement, and routing, which results in modifications to the switching activity, timing profiles, spatial electromagnetic emanations, and consequently, the observable side-channel leakage of the circuit. As a result, the probing and scanning techniques utilized to identify target points of interest on the circuit for non-invasive analysis are inherently limited as the underlying implementation of a circuit results in differences from the assumptions defined by an attack methodology.
  
\end{itemize}
\section{FPGA Device Characterization And Definition of the Scanned Region}
The physical characteristics of the target FPGA device are described in this section, along with the framework for partitioning the scanned region. The process node and the inventory of resources included on the device is described in Section 3.1. An in-depth description of the available clock resources of the target FPGA is provided in Section 3.2. Finally, the die partitioning methodology developed to probe and characterize the EM emanations of the FPGA is described in Section 3.3. The outline of the package, the device floorplan, and high resolution images of the target region are utilized to analyze the EM emanations, which define each scan point for near-field measurements.

\subsection{Overview of Target Device}

EM scans were performed on an AMD (Xilinx) Artix-7 XC7A100T-1CSG324C FPGA, which is mounted on a Digilent Nexys A7 development board, as illustrated in Fig.~\ref{fig:Nexys_A7}. The Artix-7 is fabricated in a 28 nm high-k metal-gate, high-performance low-power (HPL) CMOS process and packaged in a ball grid array (BGA) of 324 solder bumps (CSG324) each with a dimension of 15 mm × 15 mm and a pitch of 0.8 mm between bumps \cite{xilinx_ds180}.

The XC7A100T is the second-largest device in the Artix-7 series, integrating 101,440 configurable logic cells organized into 15,850 logic slices, and providing 1,188 Kb of distributed RAM. The FPGA fabric also includes 240 DSP48E1 slices and on-chip block random access memory (BRAM) comprised of 270 × 18 Kb and 135 × 36 Kb tiles. The die area is divided into eight clock regions arranged in a 2-column by 4-row grid (X0Y0-X1Y3). Six of the clock regions X0Y0, X0Y1, X0Y2, X0Y3, X1Y1, and X1Y2, include a clock-management tile (CMT), with each tile comprised of a phased-locked loop (PLL) and a mixed-mode clock manager (MMCM) \cite{xilinx_ds180}. The remaining two regions, X1Y0 and X1Y3, share the global clocking resources of the adjacent CMT in the corresponding row \cite{xilinx_ug953}.

\subsection{Xilinx 7-Series FPGA Clocking Resources}
The Artix-7 is a member of the AMD 7-series FPGA family, which are characterized by a hierarchical clocking architecture centered around clock regions that scale proportionally with device count. The smallest FPGA of the 7-series contains a single clock region, while the largest is comprised of twenty-four clock regions. As outlined in Section 3.1, the Artix-7 features eight clock regions with six clock management tiles (CMT). The global clock network includes thirty-two clock lines, driven by global clock buffers, BUFG and BUGCTRL, which facilitate glitch-free clock enable, gating, and clock multiplexing at the system level. The global clock network spans the entire FPGA and provides dedicated global routing paths to all sequential and clocking elements. Within each clock region, the horizontal global clock tree spine is driven by the horizontal clock buffers (BUFH), with a total of twelve BUFHs available within each clock region.

In addition to global clock routing resources, the 7-series FPGAs include clock management tiles (CMTs), which serve as dedicated clock conditioning modules. The phase-locked loop (PLL) and mixed-mode clock manager (MMCM) of a CMT are each driven by one or more clock inputs. The CMT modules generate multiple derived clock frequencies using a programmable divide or multiply ratio and an adjustable duty cycle of the primary clock signal. The MMCMs allow for fine and dynamic phase shifting, which enables the runtime adjustment of the phase of derived clock signals while the circuit remains in the active mode \cite{xilinx_ug472}.

\begin{figure}[t]
  \centering
  \begin{subfigure}[t]{0.2\textwidth}
    \centering    \includegraphics[width=\textwidth,height=0.65\textheight,keepaspectratio]{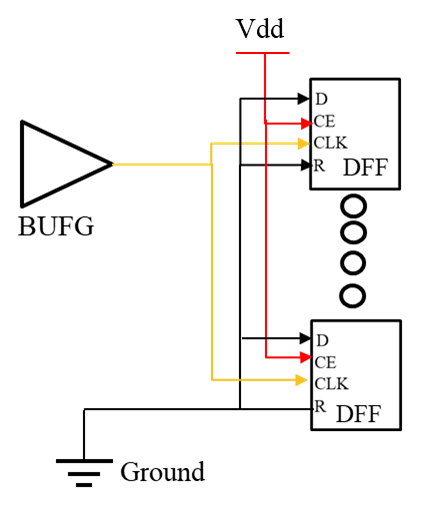}
    \caption{}
    %\caption{Config\_0: D input state tied to ground.}
    \label{fig:config0}
  \end{subfigure}\hfill
  \begin{subfigure}[t]{0.2\textwidth}
    \centering    \includegraphics[width=\textwidth,height=0.65\textheight,keepaspectratio]{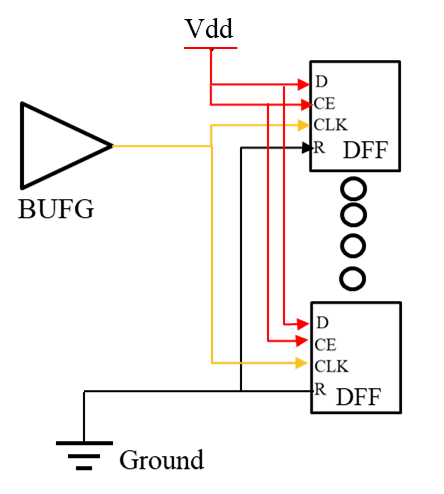}
    %\caption{Config\_1: D input state tied to supply.}
    \caption{}
    \label{fig:config1}
  \end{subfigure}
  \caption{Two configurations of a chain of DFFs in a CLB slice, where (a) Config\_0 ties the input state D of all flip-flops to ground and (b) Config\_1 ties the input state D of all flip-flops to the $V_{dd}$ supply.}
  \label{fig:full_configs}
\end{figure}

\subsection{Structure of the Near-Field Scan-Region}
The specifications of the packages of AMD (Xilinx) 7-series FPGAs vary depending on the type of bond applied between the die and the package, the pitch (mm) of the I/O pads, the size (mm) of the package, and the maximum number of I/Os. The Artix-7 is encapsulated and wire-bonded with a chip-scale CSG324 BGA package that supports up to 210 user-programmable I/O pads \cite{xilinx_ug475}.

The dimensions of the active silicon die, which includes the central logic array and the metal interconnect layers, were micro-photographed through computed tomography (CT) scans at the Idaho National Laboratory (INL). The scanned images of the target Artix-7 FPGA are shown in Fig.~\ref{fig: CT_scan}. From the CT images, the active silicon die is approximately 6.87 mm by 10.01 mm. The nominal die area (active silicon die with package redistribution), the active silicon die dimensions, and key package dimensions, including the overall length, width, ball diameter, and pitch of the BGA micro-bumps are listed in Table~\ref{tab:dimensions}. By analyzing various package variants that share the same 0.8 mm BGA pitch \cite{xilinx_ug475}, the nominal die dimensions were estimated to be 11.10 mm by 12.05 mm.

\section{Experimental Configuration And Operating Conditions}

In this section, the methodology that was developed to characterize the effect of the underlying circuit activity of the FPGA on the emanated electromagnetic (EM) fields is described. The equipment, framework, and scanning methodology utilized to investigate the near-field electromagnetic emanations produced by the target FPGA are described in Section 4.1. The operating conditions of the FPGA, which are defined to conduct a series of analyses of the emanating EM field, are described in Section 4.2.

\subsection{Measurement Framework and Scanning Methodology}
The components of the experimental EM scan station used to measure the near-field electromagnetic emissions from the target Artix-7 FPGA are shown in Fig.~\ref{fig:EM_SC_setup}. The EM scan station consists of a motion controller, an XYZ scan table, a mounted target FPGA, an H-Field probe (RFS-B 0.3-3), a probe amplifier, and a mixed signal oscilloscope with an analog bandwidth of up to 33 GHz and a sampling rate of 80 GSa/s.   
\begin{table}[t]
  \centering
  \caption{Physical dimensions of the package and IC of the Artix-7.}
  \label{tab:dimensions}
  \begin{tabular}{lcr}
    \toprule
    Parameter & Dimensions (mm) \\
    \midrule
    Ball diameter (nominal) & 0.45 \\
    BGA pitch & 0.8 \\
    Active die region & 6.87 $\times$ 10.01 \\
    Nominal die region & 11.10 $\times$ 12.05 \\
    Package length and width & 15 $\times$ 15 \\
        
    \bottomrule
  \end{tabular}
\end{table}

To efficiently scan for EM emissions, a graphical user interface (GUI) was created at INL that provides control between the target FPGA device, the motion controller of the scan table, and the oscilloscope, allowing the user to configure grid parameters for point measurements based on the dimensions listed in Table~\ref{tab:dimensions}. Utilizing a grid of 56 points in the x-dimension and 63 points in the y-dimension, the near H-field probe moves across the surface of the packaged FPGA in a serpentine fashion, collecting data at each grid point. The step size is defined in terms of either the total number of steps or absolute coordinates in micrometers, where one step corresponds to 2.5~\textmu m.

In the current configuration, the acquisition mode is set to peak-detect through the command-line provided within the graphical user interface (GUI). The peak-detect mode facilitates the detection of exceedingly narrow and rapid glitches that might be overlooked by standard sampling methods. 
When operating in peak-detect mode, the oscilloscope samples the input with a sampling rate of 80 GSa/s, but instead of exporting the full raw sample stream, time is divided into acquisition buckets for which the minimum and maximum voltage observed in each bucket is recorded. 
The header indicates a bucket interval of 1 ns, which corresponds to a bucket rate of 1 GSa/s (i.e., one bucket per nanosecond). For the capture performed, the recorded trace contains a million acquisition points, corresponding to 1 ms of data. Consequently, given a sampling rate of 80 GSa/s and the 1 ns bucket interval, approximately 80 internal samples are evaluated by the oscilloscope within each acquisition interval (bucket) to determine $V_{\mathrm{\text{min}}}$ and $V_{\mathrm{\text{max}}}$. The collected minimum and maximum values per bucket are subsequently represented as a voltage difference $V_{\mathrm{p\text{-}p}}$, as given by \eqref{eq:vpp}. Utilizing the peak-detect acquisition mode, therefore, results in one million $V_{\mathrm{p\text{-}p}}$ measurements collected at each grid point.

\begin{equation}
  V_{\mathrm{p\text{-}p}}=\left|V_{\max}-V_{\min}\right|
  \tag{1}            % forces the number (1) even if it isn't the first equation
  \label{eq:vpp}
\end{equation}

\subsection{FPGA Operating Conditions}
In 7-series AMD (Xilinx) FPGAs, each combinational logic block (CLB) consists of two slices with each slice comprised of four LUTs, a carry chain, and eight DFFs.
To evaluate the effect on clock network resources and the clock distribution network, the LUTs and carry chain were not activated during either of the two experiments performed.
\begin{figure}[t]
  \centering

  % -------- Row 1 (a-d) --------
  \begin{subfigure}[t]{0.27\columnwidth}
    \centering
    \includegraphics[width=\linewidth,height=2.5cm,keepaspectratio]{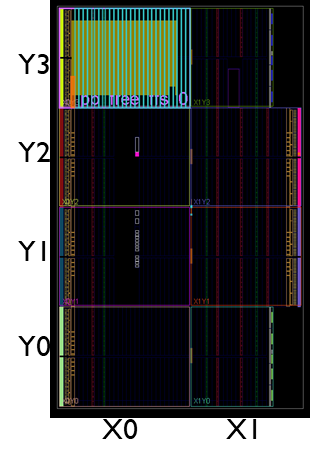}
    \caption{}
    \label{fig:lay_a}
  \end{subfigure}\hfill
  \begin{subfigure}[t]{0.24\columnwidth}
    \centering
    \includegraphics[width=\linewidth,height=2.5cm,keepaspectratio]{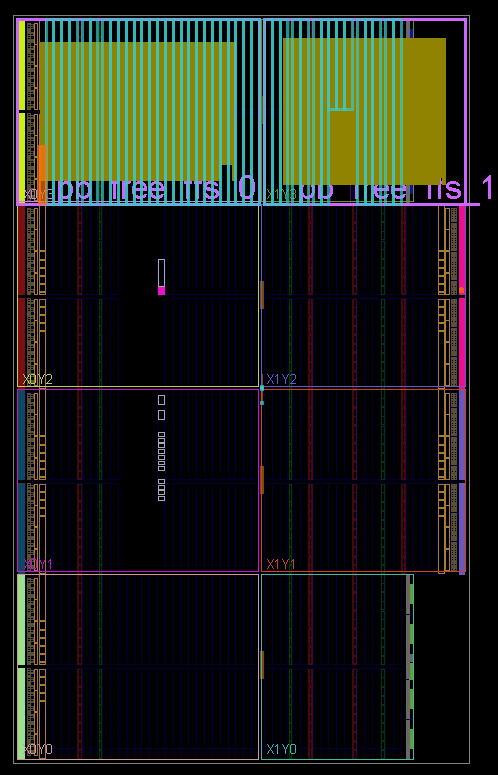}
    \caption{}
    \label{fig:lay_b}
  \end{subfigure}\hfill
  \begin{subfigure}[t]{0.24\columnwidth}
    \centering
    \includegraphics[width=\linewidth,height=2.5cm,keepaspectratio]{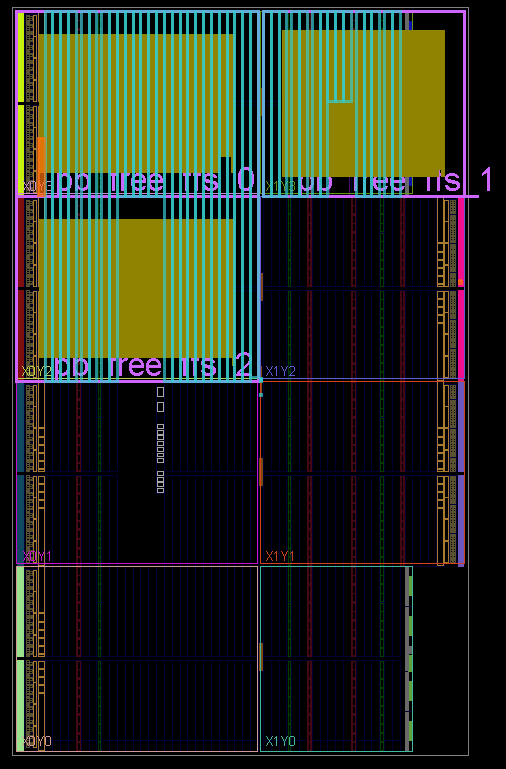}
    \caption{}
    \label{fig:lay_c}
  \end{subfigure}\hfill
  \begin{subfigure}[t]{0.24\columnwidth}
    \centering
    \includegraphics[width=\linewidth,height=2.5cm,keepaspectratio]{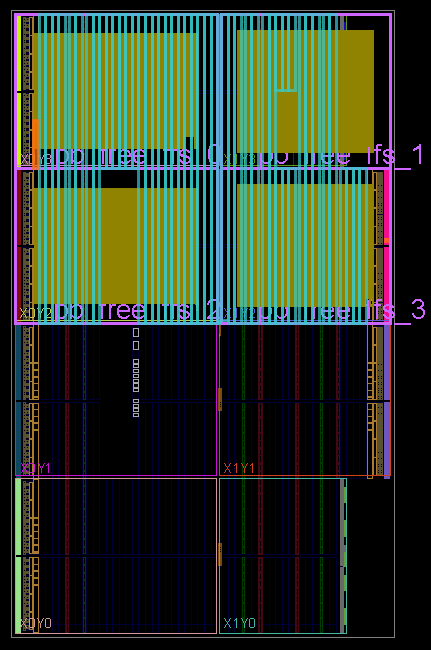}
    \caption{}
    \label{fig:lay_d}
  \end{subfigure}

  \vspace{4pt}

  % -------- Row 2 (e-h) --------
  \begin{subfigure}[t]{0.24\columnwidth}
    \centering
    \includegraphics[width=\linewidth,height=2.5cm,keepaspectratio]{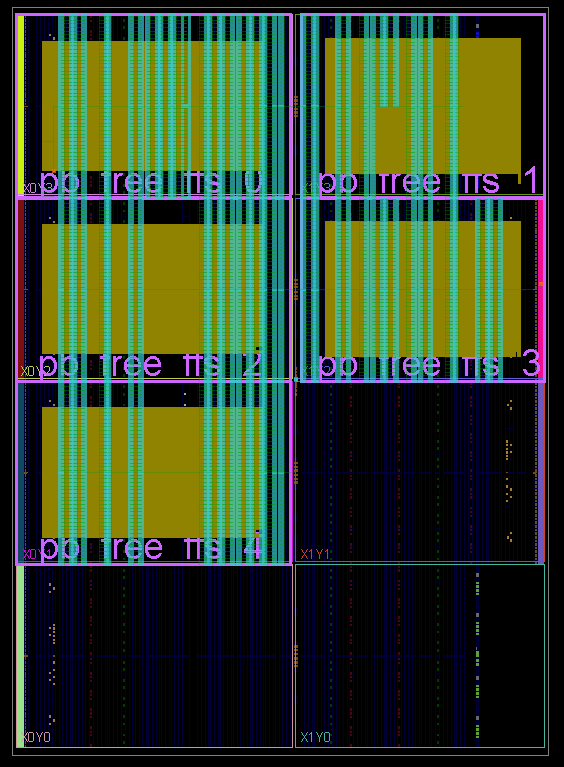}
    \caption{}
    \label{fig:lay_e}
  \end{subfigure}\hfill
  \begin{subfigure}[t]{0.24\columnwidth}
    \centering
    \includegraphics[width=\linewidth,height=2.5cm,keepaspectratio]{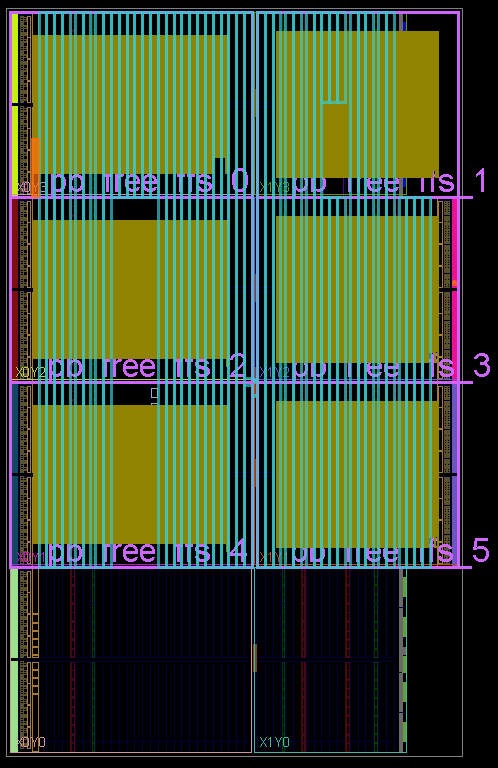}
    \caption{}
    \label{fig:lay_f}
  \end{subfigure}\hfill
  \begin{subfigure}[t]{0.24\columnwidth}
    \centering
    \includegraphics[width=\linewidth,height=2.5cm,keepaspectratio]{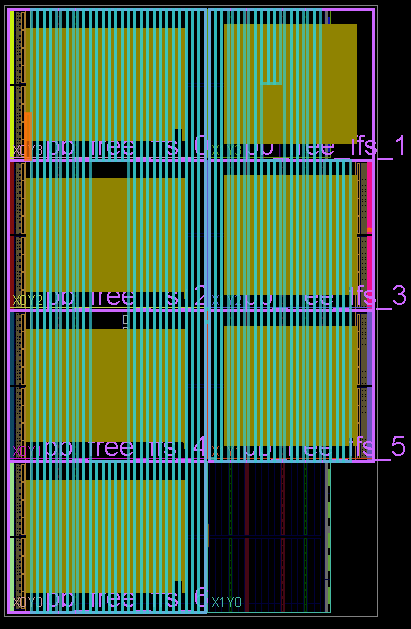}
    \caption{}
    \label{fig:lay_g}
  \end{subfigure}\hfill
  \begin{subfigure}[t]{0.24\columnwidth}
    \centering
    \includegraphics[width=\linewidth,height=2.5cm,keepaspectratio]{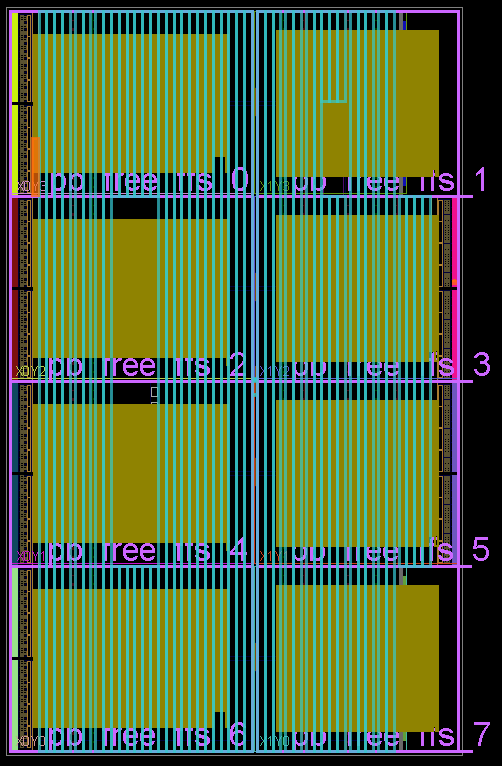}
    \caption{}
    \label{fig:lay_h}
  \end{subfigure}

  \caption{Layout of increasing active DFFs within clock regions. (a) Baseline: One clock region active, (b) Two clock regions active, (c) Three clock regions active, (d) Four clock regions active, (e) Five clock regions active, (f) Six clock regions active, (g) Seven clock regions active, and (h) Eight clock regions active}
  \label{fig:Incr_CR}
\end{figure}

Two sets of experiments were conducted to characterize the effect of device placement, frequency, and the amount of active clocking resources on the EM emanations generated by the clock distribution network.
In the first set of experiments, which evaluate the effect of frequency on EM emanations, two distinct configurations were defined, with both providing characterization of EM emanation at operating frequenies of 100 MHz and 300 MHz. For the first configuration, config\_0, all state inputs and resets were tied to ground, clock enables were connected to the supply voltage, and the clock inputs were connected to an MMCM through the global clock buffers (BUFG), as illustrated in Fig.~\ref{fig:config0}. For the second configuration, config\_1, all state inputs and clock enables were connected to the $V_{\mathrm{d\text{d}}}$ supply voltage, resets were grounded, and the clock inputs were connected to an MMCM through the global clock buffers (BUFG), as depicted in Fig.~\ref{fig:config1}.

In the second set of experiments, the configuration of the clock network and the clock frequency were fixed while the load on the clock network was increased incrementally by activating DFFs from the eight clock regions, as shown in Fig.~\ref{fig:Incr_CR}. For the second set of experiments, the second configuration was utilized, where all input states and clock enables were connected to $V_{\mathrm{d\text{d}}}$. In addition, the clocking frequency was set to 200 MHz.

\section{Experimental Results And Discussion}
In this section, the results from the analysis of the scanned EM profile of the Artix-7 FPGA for the two sets of experiments described in Section 4.2 are presented and discussed.
As described in Section 4.1, for the scans of the near field EM, each scanned grid location captures a million $V_{\mathrm{p\text{-}p}}$ samples using the experimental setup shown in Fig.~\ref{fig:Complete_EM_Setup}. Traces are collected of the EM field through the peak detect mode of the MSOV334A oscilloscope, which captures narrow transients and rare peaks.

The $V_{\mathrm{p\text{-}p}}$ is the primary metric utilized to directly represent the magnitude of the observed peak-to-peak swings in the waveform of the H-field. The initial results from the scanned Artix-7 represent the maximum $V_{\mathrm{p\text{-}p}}$ observed at each grid location across one million sampled points. The determined $V_{\mathrm{p\text{-}p}}$ does not provide a complete analysis of the underlying circuit behavior as changes in placement, frequency, or utilized clock resources effect the emitted EM field, even if a given grid location between two scans results in a difference in the calculated maximum $V_{\mathrm{p\text{-}p}}$ value. The $V_{\mathrm{p\text{-}p},max}$ per grid location does not distinguish between grid points that are consistently active and grid points that are usually inactive but occasionally produce large transients.

To derive meaningful interpretations from the collected samples, the mean and variance are computed to quantify the typical peak-to-peak variation at each scanned location and the change in EM emanations over time. Both mean and variance provide more detailed analysis of changes in EM field emanations when utilizing peak detect mode rather than relying on extrema only for each bucket of data points. The mean and variance at each grid location for a complete scan of 1 million points are computed using \eqref{eq:mean} and \eqref{eq:variance}, respectively.
\begin{figure}[t]
  \centering
  \begin{subfigure}[t]{\columnwidth}
    \centering
    \includegraphics[width=0.65\linewidth]{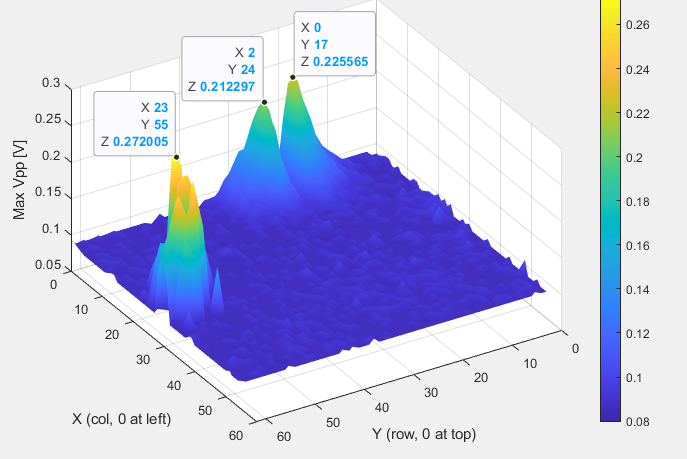}
    %\caption{For operating Frequency 100MHz.}
    \caption{}
    \label{fig:img1_stack}
  \end{subfigure}

  \vspace{2pt}

  \begin{subfigure}[t]{\columnwidth}
    \centering
    \includegraphics[width=0.65\linewidth]{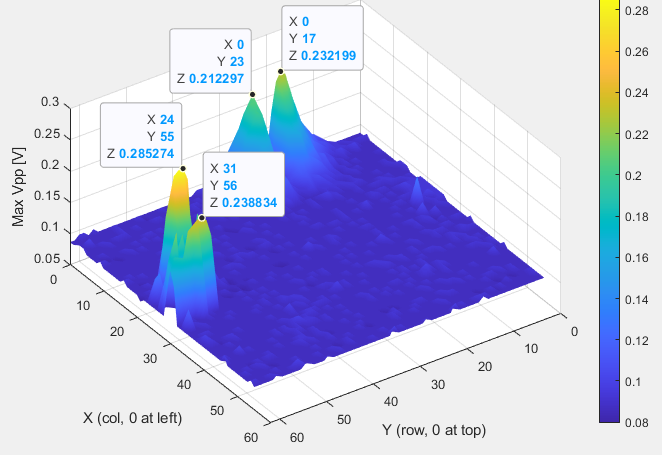}
    %\caption{For operating Frequency 300MHz.}
    \caption{}
    \label{fig:img2_stac}
  \end{subfigure}
  \caption{$V_{\mathrm{p\text{-}p},max}$ map for state tied to ground at an operating frequency of (a) 100 MHz and (b) 300 MHz.}
  \label{fig:Congif_0_maps}
\end{figure}

To further assess variations in EM emanations due to frequency, DFF placement, and number of utilized clocking resources, mean-difference and variance-difference maps are generated by comparing the mean and variance maps of two configurations of the Artix-7 FPGA on a grid-point-by-grid-point basis. The difference in mean $\Delta\mu(x,y)$ and the difference in variance $\Delta\sigma^2(x,y)$ between two maps are computed using ~\eqref{eq:mean_difference} and ~\eqref{eq:variance_difference}, respectively. The calculated differences help identify statistically significant changes in underlying circuit behavior.
  
\begin{equation}
\mu(x,y)=\frac{1}{N}\sum_{k=1}^{N} V_{pp}[k]
\tag{2}
\label{eq:mean}
\end{equation}

\begin{equation}
\sigma^{2}(x,y)=\frac{1}{N-1}\sum_{k=1}^{N}\bigl(V_{pp}[k]-\mu(x,y)\bigr)^{2}
\tag{3}
\label{eq:variance}
\end{equation}

\begin{equation}
\Delta\mu(x,y)=\mu_{A}(x,y)-\mu_{B}(x,y).
\tag{4}
\label{eq:mean_difference}
\end{equation}

\begin{equation}
\Delta\sigma^{2}(x,y)=\sigma^{2}_{A}(x,y)-\sigma^{2}_{B}(x,y).
\tag{5}
\label{eq:variance_difference}
\end{equation}

\begin{table}[h!]
\centering
\small
\caption{Analysis of the mean and variance of the EM emanations for Config\_0 at 100 MHz and 300 MHz.}
\label{tab:experiment_1}

% -------- Table 1 --------
\setlength{\tabcolsep}{3pt}
\renewcommand{\arraystretch}{1.02}

\sisetup{
  output-exponent-marker = \mathrm{e},
  exponent-product = {},
  table-align-exponent = false,
  table-number-alignment = center,
  group-digits = false
}

\begin{tabular}{@{} l
    S[table-format=-1.8e-2, table-column-width=1.55cm]
    S[table-format=-1.8e-2, table-column-width=1.55cm]
    S[table-format=-1.8e-2, table-column-width=1.55cm] @{}}

\toprule
\textbf{Metric}
& \multicolumn{1}{c}{\textbf{Min}}
& \multicolumn{1}{c}{\textbf{Med}}
& \multicolumn{1}{c}{\textbf{Max}} \\
\midrule

Mean $\mu$ (100) & 0.0465534    & 0.0469709    & 0.0845099 \\
Mean $\mu$ (300) & 0.0463575    & 0.0467533    & 0.0846009 \\
Var $\sigma^2$ (100)  & 5.04948e-05  & 5.12232e-05  & 0.00170133 \\
Var $\sigma^2$ (300)  & 5.01959e-05  & 5.09993e-05  & 0.00172461 \\

\midrule

$\Delta\mu$ & -0.00195823   & 0.000215561  & 0.00141386 \\
$\Delta\sigma^2$  & -2.86048e-05  & 2.30765e-07  & 1.59673e-05 \\

\bottomrule
\end{tabular}

\end{table}

\begin{figure}[t]
  \centering
  \includegraphics[width=0.7\columnwidth]{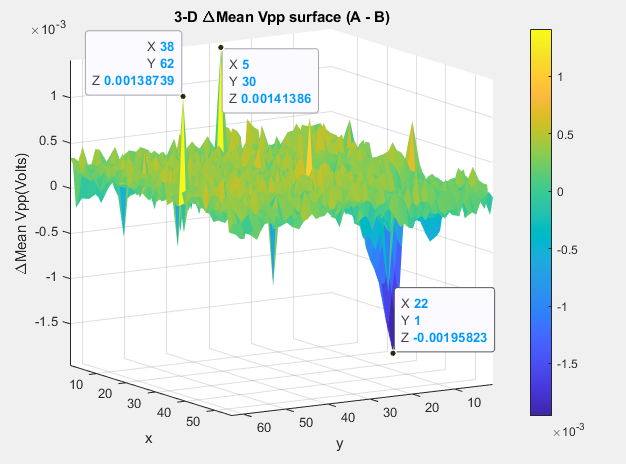}
  \caption{Calculated $\Delta\mu(x,y)$ when the input is tied to ground. $\Delta\mu(x,y)$ determined between FPGA programmed for 100 MHz and 300 MHz operation.}
  \label{fig:del_conf_0}
\end{figure}

In the first set of experiments, two configurations of the FPGA, as described in Section 4.2 and illustrated in Fig.~\ref{fig:full_configs}, are utilized to evaluate the effect of frequency on EM emanation, with the same circuit and placement implemented on the Artix-7. The $V_{\mathrm{p\text{-}p},max}$ is determined and plotted as shown in Fig.~\ref{fig:Congif_0_maps} for config\_0 where the input is tied to ground. The EM emanations are analyzed at operating frequencies of 100 MHz and 300 MHz. The difference in the mean $V_{\mathrm{p\text{-}p}}$ for config\_0 between a circuit operating at 100 MHz and 300 MHz is shown in Fig.~\ref{fig:del_conf_0}.  

\begin{table}[h!]
\centering
\small
\caption{Analysis of the mean and variance of the EM emanations for Config\_1 at 100 MHz and 300 MHz.}
\label{tab:experiment_2}

% -------- Table 1 --------
\setlength{\tabcolsep}{3.5pt}
\renewcommand{\arraystretch}{1.02}

\sisetup{
  output-exponent-marker = \mathrm{e},
  exponent-product = {},
  table-align-exponent = false,
  table-number-alignment = center,
  group-digits = false
}
\begin{tabular}{@{} l
    S[table-format=-1.8e-2, table-column-width=1.55cm]
    S[table-format=-1.8e-2, table-column-width=1.55cm]
    S[table-format=-1.8e-2, table-column-width=1.55cm] @{}}
\toprule
\textbf{Metric} & {\textbf{Min}} & {\textbf{Med}} & {\textbf{Max}} \\
\midrule
Mean $\mu$ (100) & 0.0464825   & 0.0468474   & 0.0851531 \\
Mean $\mu$ (300) & 0.0464187   & 0.0467834   & 0.0853271 \\
Var $\sigma^2$ (100)  & 5.03881e-05 & 5.11618e-05 & 0.00173164 \\
Var $\sigma^2$ (300)  & 5.01796e-05 & 5.09824e-05 & 0.00175918 \\
\midrule
$\Delta\mu$ & -0.00207057 & 0.000293998 & 0.00100211 \\
$\Delta\sigma^2$  & -4.79505e-05 & 1.68226e-07 & 2.13343e-05 \\
\bottomrule
\end{tabular}

\end{table}

\begin{figure}[h!]
  \centering
  \includegraphics[width=0.7\columnwidth]{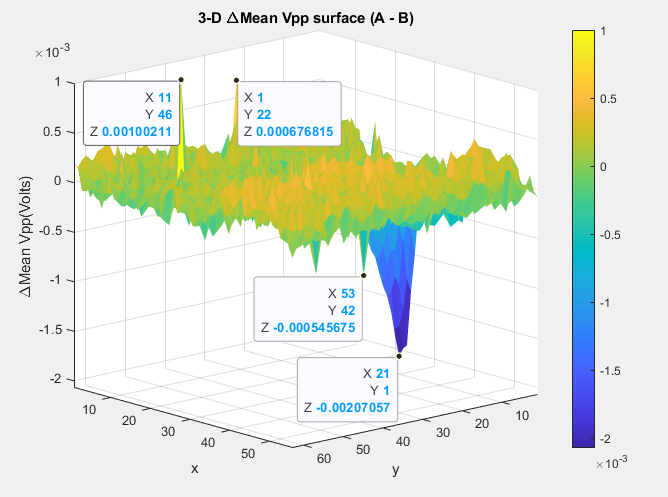}
  \caption{Calculated $\Delta\mu(x,y)$ when the input is tied to $V_{\mathrm{\text{dd}}}$. $\Delta\mu(x,y)$ determined between FPGA programmed for 100 MHz and 300 MHz operation.}
  \label{fig:del_conf_1}
\end{figure}

\begin{table*}[t]
\centering
\small
\setlength{\tabcolsep}{5pt}
\renewcommand{\arraystretch}{1.05}
\caption{Summary of results for Config\_1 when the number of utilized clock regions is varied from 2 to 8.}
\label{tab:Config1_diff_util}

\begin{tabularx}{\textwidth}{@{} >{\raggedright\arraybackslash}p{0.14\textwidth} *{10}{Y} @{}}
\toprule
\multirow{2}{*}{\makecell[l]{\textbf{Number of}\\\textbf{Clock Regions}}}  &
\multicolumn{2}{c}{\textbf{Number of Active DFFs}} &
\multicolumn{2}{c}{\textbf{Mean-Difference ($V$)}} &
\multicolumn{2}{c}{\textbf{Variance-Difference ($V^2$)}} &
\multicolumn{2}{c}{\makecell[l]{\textbf{$\Delta\mu$ Flagged Grid }\\\textbf{points}}} &
\multicolumn{2}{c}{\makecell[l]{\textbf{$\Delta\sigma^{2}$ Flagged Grid}\\\textbf{points}}} \\
\cmidrule(lr){2-3}\cmidrule(lr){4-5}\cmidrule(lr){6-7}\cmidrule(lr){8-9}\cmidrule(lr){10-11}
& \textbf{Baseline Design} & \textbf{Incremental Design}
& \textbf{Minimum} & \textbf{Maximum}
& \textbf{Minimum} & \textbf{Maximum}
& \textbf{Threshold} & \textbf{Number of grid points}
& \textbf{Threshold} & \textbf{Number of grid points} \\
\midrule
2 & 20,800 & 31,600 & -0.0032 & 0.003297 & \mbox{-3.329e-05} & \mbox{6.397e-05} & 0.000314 & 568/3528 & \mbox{7.882e-07} & 585/3528 \\
3 & 20,800 & 47,570 & -0.0028 & 0.0033 & \mbox{-5.093e-05} & \mbox{3.234e-05} & 0.00034 & 603/3528 & \mbox{7.940e-07} & 627/3528 \\
4 & 20,800 & 62,740 & -0.0018 & 0.0035 & \mbox{-2.153e-05} & \mbox{6.455e-05} & 0.00037 & 596/3528 & \mbox{9.425e-07} & 609/3528 \\
5 & 20,800 & 78,710 & -0.0020 & 0.0032 & \mbox{-3.272e-05} & \mbox{4.046e-05} & 0.00052 & 548/3528 & \mbox{1.056e-06} & 677/3528 \\
6 & 20,800 & 93,880 & -0.0033 & 0.0032 & \mbox{-5.458e-05} & \mbox{7.643e-05} & 0.00049 & 659/3528 & \mbox{1.068e-06} & 760/3528 \\
7 & 20,800 & 114,680 & -0.0061 & 0.0030 & \mbox{-5.788e-05} & \mbox{9.103e-05} & 0.00048 & 753/3528 & \mbox{1.103e-06} & 817/3528 \\
8 & 20,800 & 126,480 & -0.0022 & 0.0033 & \mbox{-7.008e-05} & \mbox{1.080e-04} & 0.00061 & 613/3528 & \mbox{1.185e-06} & 723/3528 \\
\bottomrule
\end{tabularx}
\end{table*}
 A positive $\Delta\mu$ value at a given (x,y) location indicates that, at the same spatial location, the circuit operating at 100 MHz exhibits a larger average peak-to-peak voltage $V_{\mathrm{p\text{-}p}}$ than the same circuit operating at 300 MHz. A positive $\Delta\mu(x,y)$, therefore, indicates that the H-field for the circuit operating at 100 MHz produces a larger typical $V_{\mathrm{p\text{-}p}}$ measurement at the given (x,y) location. 
Conversely, a negative $\Delta\mu(x,y)$ value indicates that the circuit operating at 300 MHz results in a larger peak-to-peak voltage $V_{\mathrm{p\text{-}p}}$ measurement at the given (x,y) grid location. Grid points with near-zero difference imply that the typical measured $V_{\mathrm{p\text{-}p}}$ is similar at both 100 MHz and 300 MHz circuit operation.  
 
Metrics computed for circuits implemented with config\_0 are listed in Table~\ref{tab:experiment_1} for identical circuit placements and for operating frequencies of 100 MHz and 300 MHz. For the analysis of variance-difference, a positive $\Delta\sigma^{2}(x,y)$ indicates higher variability in the EM emanations, implying more bursty or intermittent $V_{\mathrm{p\text{-}p}}$ measurements in the circuit implemented with config\_0 operating at 100 MHz than the circuit implemented with config\_0 operating at 300 MHz.

Conversely, a negative $\Delta\sigma^{2}(x,y)$ indicates a higher variability in the circuit implemented with config\_0 operating at 300 MHz than in the circuit implemented with config\_0 operating at 100 MHz. To determine the number of grid points whose values deviate from the typical calculated mean difference and variance difference, the median absolute deviation (MAD) criterion is utilized to define a threshold for the computed differences. First, the grid-typical value $m$ of a mean or variance difference map M(x,y) is calculated by taking the median of the mean difference $\Delta\mu(x,y)$  or variance difference $\Delta\sigma^{2}(x,y)$ across all grid locations ($x$, $y$), as given by
\begin{equation}
m=\mathrm{median}\!\left(M(x,y)\right).
\tag{6}
\label{eq:grid_median}
\end{equation}
Next, the absolute deviation map $D(x,y)$ is determined by subtracting the calculated median of the entire grid $m$ from the mean-difference or variance-difference at each grid point $M(x,y)$ and taking the absolute value of the difference, as given by
\begin{equation}
D(x,y)=\left|M(x,y)-m\right|.
\tag{7}
\label{eq:abs_deviation}
\end{equation}
Finally, the MAD is computed as the median of the absolute deviations over the entire grid, as described by
\begin{equation}
\mathrm{MAD}=\mathrm{median}\!\left(D(x,y)\right).
\tag{8}
\label{eq:mad}
\end{equation}

\begin{table}[h!]
  \centering
  \caption{Analysis of flagged grid points for Config\_0 at 100 MHz and 300 MHz using the MAD Criterion.}
  \label{tab:Thres_config0_diff_freq}
  \begin{tabular}{lcr}
    \toprule
    Difference metric & Threshold & Flagged grid points \\
    \midrule
    $\Delta\mu$  & 2.94353e$-$04 & 314/3528 (8.900\%) \\
    $\Delta\sigma^2$  & 6.64209e$-$07 & 391/3528 (11.083\%) \\
        
    \bottomrule
  \end{tabular}
\end{table}

\begin{table}[h!]
  \centering
  \caption{Analysis of flagged grid points for Config\_1 at 100 MHz and 300 MHz using the MAD Criterion.}
  \label{tab:Thres_config1_diff_freq}
  \begin{tabular}{lcr}
    \toprule
    Difference metric & Threshold & Flagged grid points \\
    \midrule
    $\Delta\mu$ & 3.17997e$-$04 & 290/3528 (8.220\%) \\
    $\Delta\sigma^2$  & 6.88515e$-$07 & 434/3528 (12.302\%) \\
        
    \bottomrule
  \end{tabular}
\end{table}

%\noindent where M(x,y) denotes the selected difference map, either the mean-difference $\Delta\mu$ or variance-difference $\Delta\sigma^{2}$, computed from the 1,000,000 peak detect $V_{\mathrm{p\text{-}p}}$ samples at each grid point of two configurations of the FPGA. 
Grid points are flagged as hotspots when $D(x,y)$ exceeds three times the determined MAD value for the corresponding difference map ~\cite{mad}.  
As listed in Table~\ref{tab:Thres_config0_diff_freq}, a total of 314 grid points are flagged through the analysis of the mean-difference, which indicates a significant shift from the average measured $V_{\mathrm{p\text{-}p}}$ or a large shift in the peak-to-peak swing on grid points of the circuit configured using config\_0 operating at 100 MHz and config\_0 operating at 300 MHz. 
In addition, a total of 391 grid points are identified through the analysis of the variance-difference. Furthermore, a total of 96 grid points are identified to be common in both the calculated mean difference and variance difference, indicating locations where the temporal variability of $V_{\mathrm{p\text{-}p}}$ exhibits substantial divergence, which suggests a change in intermittency or burstiness rather than a simple shift in DC level.

For the second configuration of experiment 1, config\_1 defines the state of the DFFs of each clocking region, for which all state inputs are tied to the $V_{\mathrm{d\text{d}}}$ supply. A similar analysis was performed for circuits considered under config\_1, where grid points flagged using the MAD criterion on the difference metrics are identified. The mean-difference map is shown in Fig.~\ref{fig:del_conf_1}.
Metrics computed for circuits implemented with config\_1 for the same placement and configuration of devices on the FPGA but with 100 MHz and 300 MHz operating frequency are listed in Table~\ref{tab:experiment_2}. 
The minimum value (most negative) of the mean-difference implies that for the given grid point, the circuit operating at 300 MHz exceeds the typical measured $V_{\mathrm{p\text{-}p}}$ by the largest amount as compared to the same circuit operating at 100 MHz. Similarly, the maximum value (most positive) of the mean-difference implies that for the given grid point, the circuit operating at 100 MHz exceeds the typical peak-to-peak voltage $V_{\mathrm{p\text{-}p}}$ by the largest amount as compared to the same circuit operating at 300 MHz. 
As listed in Table~\ref{tab:Thres_config1_diff_freq}, a total of 290 grid points exceed the threshold set by the MAD criterion when applied to the analysis of the mean-difference. Similarly, a total of 434 grid points exceed the threshold when applied to the analysis of the variance-difference, and a total of 104 grid points are identified to be common in both the calculated differences, indicating locations where the temporal variability of $V_{\mathrm{p\text{-}p}}$ exhibits substantial divergence.

In the second set of experiments, the state configuration of the flip-flops and the operating frequency of the circuit are fixed to config\_1 and 200 MHz, respectively. 
The second experiment provides an analysis of the effect on the emanating EM fields due to an incremental increase in the number of clocked elements, specifically DFFs, and the utilization of clock regions, which are shown in Fig.~\ref{fig:Incr_CR}. 
The baseline circuit configuration implemented on the FPGA for experiment 2 is the activation of all clocked elements (i.e., DFFs) within the clock region X0Y3. 
The comparative analysis is performed by juxtaposing the implemented baseline circuit with circuits that activate a varying number of clock regions of the FPGA, with all DFFs in a given clock region also activated. To facilitate analysis, the data listed in Table~\ref{tab:Config1_diff_util} includes the number of clocked elements, the number of clock regions, the mean-difference metric, the variance-difference metric, and the number of flagged grid points.

The number of activated DFFs increases based on the total number of DFFs available in any given clock region. For the results listed in Table~\ref{tab:Config1_diff_util}, nearly 99.75\% of all available DFFs within the core fabric of the Artix-7 are activated when all eight clock regions are utilized. In general, the number of flagged grid points increases with increasing utilization of FPGA clock regions, which indicates that a large portion of grid locations are affected. However, in some instances when the number of DFFs was increased, a slight decrease in the number of flagged grid points was observed. 
The reduction in the identified points is primarily attributed to the threshold used to flag grid points, where the threshold is set to three times the computed MAD criterion. Since the threshold is data-driven, the added clock resources introduce greater variability in background circuit activity, which consequently renders the threshold more conservative. As a result, moderate differences in $\Delta\mu$ and $\Delta\sigma^{2}$ do not exceed the specified threshold level. 

\section{Conclusion}

A novel approach is introduced to analyze the electromagnetic (EM) side-channel emanations of programmable hardware including field-programmable gate arrays (FPGAs). Instead of utilizing a fixed architecture for a specific algorithm, the nature of the underlying circuit implemented on the FPGA is defined through the programming of deterministic finite state machines (DFSMs). Any given DFSM is comprised of a control path and a data path, with the clock functioning as a shared element that governs the process flow of data and control signals. Accordingly, this study characterizes the impact of clocking resource utilization on the EM side-channel emanations of an FPGA while considering the placement, frequency, and quantity of resources employed. Statistical analyses of the mean difference and variance difference were utilized to evaluate the impact that FPGA operating frequency has on EM emanations, where both the placement and the number of resources were held constant. Approximately 8\% to 9\% of the grid points were flagged after performing an analysis of the mean difference, while approximately 11\% to 13\% were flagged after performing an analysis of the variance difference.  For the analysis of the impact of physical placement and resource utilization when operating the FPGA at 200 MHz, the number of flagged grid points increased with an increasing  number of allocated resources. A 1.32x increase in flagged grid points was observed when performing mean difference analysis between an FPGA configured to use 25\% of clock resources and the same FPGA reconfigured to use 90\% of clock resources. Similarly, a 1.39x increase in flagged grid points was observed when analyzing the variance difference for the same increase in the utilized clock resources.      

\begin{acks}
This research is part of an ongoing collaboration between Idaho National Laboratory (INL) and Drexel University to explore applications of side channel analysis of FPGA based devices to enhance protection of process control systems from cybersecurity threats. Funding for this collaboration is provided through the INL Laboratory Directed Research \& Development (LDRD) Program under DOE Idaho Operations Office Contract DE-AC07-05ID14517. This paper is marked as having identifier INL/CON-26-90537.
\end{acks}

\bibliographystyle{ACM-Reference-Format}
\bibliography{references}

\newpage

\end{document}